\documentclass[fleqn,twoside]{article}
\usepackage{espcrc2}

\usepackage{graphicx}
\usepackage{epsfig}
\usepackage[figuresright]{rotating}

\newcommand{\AmS}{{\protect\the\textfont2
  A\kern-.1667em\lower.5ex\hbox{M}\kern-.125emS}}

\title{A scintillator tile-fiber preshower detector for the CDF central calorimeter}

\author{Stefano Lami\address{\vspace*{-.1in}INFN Pisa, I-56127 Pisa, Italy}\thanks{for the CDF Collaboration}}
       
\begin{document}
\begin{abstract}
The front face of the CDF central calorimeter is being equipped with a
new Preshower detector, based on scintillator tiles
read out by WLS fibers.
%in order to cope with the luminosity increase provided by the Main Injector
%during the Tevatron's Run II data taking.
A light yield of about 40 pe/MIP at the tile exit
was obtained, exceeding the design requirements.
\vspace{-1pc}
\end{abstract}
% typeset front matter (including abstract)
\maketitle
\section{INTRODUCTION}
The physics program at the Fermilab Tevatron Collider
will continue to explore the high energy frontier of particle
physics until the commissioning of the LHC at CERN.
The luminosity increase provided by the Main Injector
requires upgrades beyond those implemented
for the first stage of the Tevatron's Run II
physics program.

The upgrade of the CDF calorimetry includes:
%\cite{NIM}: 
~(i) the replacement during
the Tevatron shutdown in Fall 2004 of the
slow gas detectors -  {\it Central Preshower} (CPR) and {\it Central Crack} (CCR)
detectors - with a faster
scintillator version which has a better segmentation, and ~(ii) the addition
of timing information to both the Central and EndPlug electromagnetic
calorimeters to filter out cosmic ray and beam related backgrounds.

This contribution focuses on the new Preshower
detector, based on 20 mm thick scintillator tiles, with a finer
segmentation than the calorimeter towers, read out by a wavelength-shifting
(WLS) fiber loaded
into a groove on the surface of each tile. It also discusses briefly the replacement
of the CCR with 5 mm thick scintillator
tiles (read out with the same technique) behind a tungsten
bar in order to cover the uninstrumented azimuthal regions of the
central calorimeter.
Each WLS fiber is spliced to a clear fiber after exiting the tile.
To optimize the final design parameters we performed several
tests comparing different scintillators, fibers and 
groove shapes.

\begin{figure}[htb]
\vspace*{-.3in}
\centering{
\epsfig{file=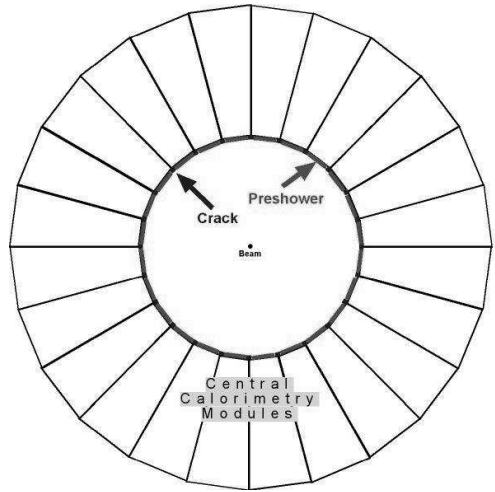,angle=0,width=15.5pc}}
\vspace*{-.36in}
\caption{The central calorimetry wedges and the
location of the Preshower and Crack detectors.}
\vspace*{-.2in}
\label{fig:calo_xsec}
\end{figure}

The location of the new detectors
on the front face of the central calorimeter, where they will
sample early showers and cover the $\phi$-cracks
between calorimeter wedges, is shown in Fig.~\ref{fig:calo_xsec}.
The CPR uses the solenoid coil and the tracking material as a radiator.

The upgrade of the CDF Central Calorimeter is expected to
play important roles in Run II physics, such as soft
electron tagging of b-jets, photon identification (ID) in SUSY
events or other new physics, and improving jet energy
resolution.

The CPR has been extensively used in electron ID,
providing about a factor 2-3 more rejection of charged pions that pass
all other cuts. This extra rejection has been crucial
in soft electron ID for b-jet tagging, as was
shown in the first {\it top} evidence paper \cite{top_evid}.
The CPR has been used in several publications involving
photon ID. By using conversion rates, which are energy independent,
it extended the QCD measurement of direct photons by more
than 100 GeV in photon transverse momentum $P_T$ \cite{photons}.

\begin{figure}[htb]
\vspace*{-.3in}
\hspace*{-.03in}\epsfig{file=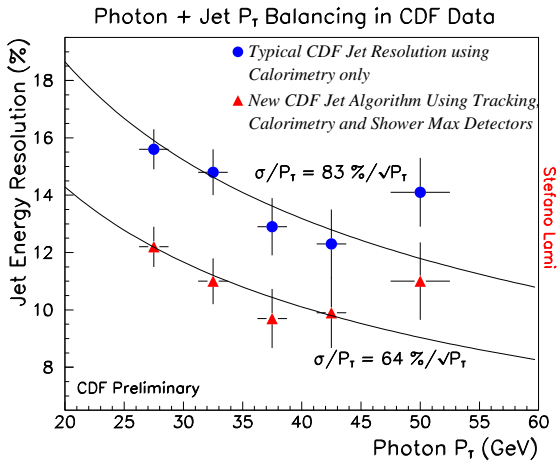,angle=0,width=19pc}
\vspace*{-.6in}
\caption{Jet energy resolution as a function of $P_T^{\gamma}$ for the standard
and PFA methods.}
\vspace*{-.15in}
\label{fig:jet}
\end{figure}

The new CPR will also be used
to improve the jet energy resolution by both correcting for energy loss in the
dead material in front of it and adding its information in jet algorithms
incorporating charged tracking.
A {\it Particle Flow Algorithm} (PFA) that combines calorimetric information with
tracking and shower max detector data was tested on a Run I photon-jet
data sample and provided a 20\% improvement in jet energy resolution~\cite{calor2000}
(see Fig.~\ref{fig:jet}).

The CCR, located after 10 radiation length thick tungsten bars,
has been checked for large pulse heights in all the rare events CDF
has observed in Run I. The $\phi$-cracks cover about 8\% of the central
detector, and in events with multiple electromagnetic objects, the 
possibility of one object hitting the crack is substantial.

%This upgrade offers an opportunity to build a much better detector
%for a relatively small cost, as the existing electronics will be reused,
%thus removing one of the largest components of the cost for any new detector
%system.

\vspace*{-.05in}
\section{DETECTOR DESIGN}

\begin{figure}[htb]
\vspace*{-.3in}
\epsfig{file=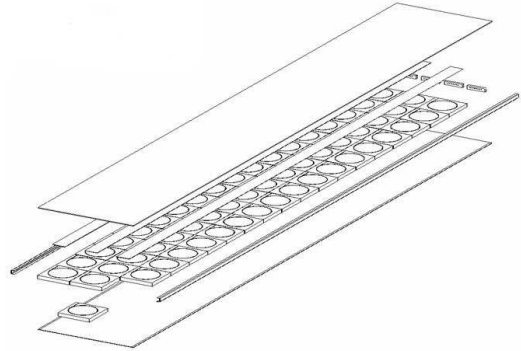,angle=0,width=16.5pc}
\vspace*{-.36in}
\caption{View of the CPR upgrade design.}
\label{fig:CPR2}
\vspace*{-.2in}
\end{figure}

The new CPR will be based on 2 cm thick scintillator tiles
segmented in $\eta$ and $\phi$ and read out by a 1 mm diameter WLS fiber
located inside a groove on the surface of each tile.
Six tiles (12.5x12.5 cm$^2$ each) will cover the front face of
each calorimeter tower, assembled
in 48 modules (Al shells) as shown in Fig.~\ref{fig:CPR2}, covering
the 48 central calorimeter wedges.
On exiting the tiles,
the WLS fibers will be spliced to clear fibers, which
will terminate into plastic connectors at the
higher $\eta$ edge of each module. There,  optical cables $\sim$5 m long will 
transmit the light to 16-channel Hamamatsu R5900
{\it PhotoMultiplier Tubes} (PMTs) at the back of the wedge.

The new CCR will use the same technique but the available space limits the scintillator
thickness to 5 mm. Ten tiles, $\sim$5 cm wide, will cover each $\phi$-crack with the
same calorimeter segmentation of 10 towers/wedge.

\begin{figure}[htb]
\vspace*{-.2in}
\epsfig{file=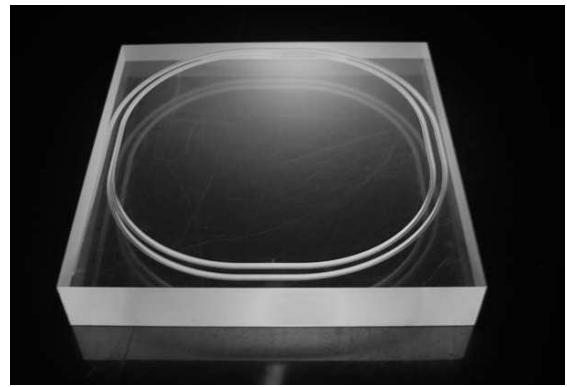,angle=0,width=17.5pc}
\vspace*{-.3in}
\caption{A CPR tile with a spiral groove path.}
\label{fig:tile}
\vspace*{-.2in}
\end{figure}

Figure~\ref{fig:tile} shows a CPR tile with a spiral groove path
for the WLS fiber. To see minimum ionizing particles (MIPs) in the CPR,
we request at least 5 photoelectrons (pe) at the PMT, which
means at least 12 pe at the tile exit, after taking in account a $\sim$60\% signal loss 
 due to light transmission by fiber splicing and optical connector
plus the clear fiber attenuation length.

\vspace*{-.15in}
\section{R$\&$D RESULTS}
%
%The italian groups (Pisa, Rome, Siena and Trieste) participating in the project
%provided a major contribution to the tile/fiber system R$\&$D program.
A cosmic ray test of individual tiles inside a light-sealed box was performed
in order to compare different scintillators (Bicron 408 and scintillator tiles provided by the
CDF JINR Dubna group), multiclad WLS and clear fibers (Kuraray and Pol.Hi.Tech),
reflectors (Al foil, Tyvek paper, 3M VM2002 reflector),
and to optimize the design of the groove in the tile.

We studied the optical coupling between the fiber and the groove,
and obtained similar results with either Bicron BC-600 glue or BC-630 grease,
about 40\% more light yield than just air coupling in a configuration
with a groove of square cross section and several loops of WLS fiber.
%The more flexible BC-630 optical grease was used in most of the tests as a 
%substitute for the BC-600 optical cement.
%
\begin{figure}[htb]
\vspace*{-.2in}
\epsfig{file=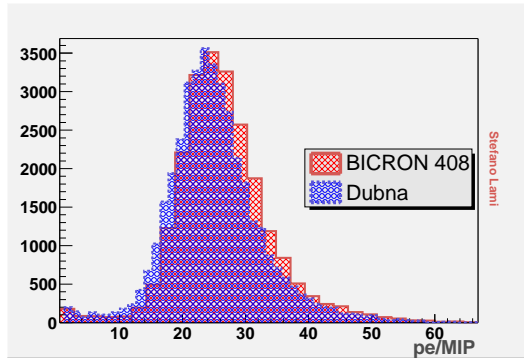,angle=0,width=16.5pc}
\vspace*{-.35in}
\caption{Light yield for two scintillators from a cosmic ray test,
with 4 loops of non-mirrored Pol.Hi.Tech fiber with grease and Tyvek.}
\label{fig:dub}
\vspace*{-.22in}
\end{figure}

We compared Dubna and Bicron 408 scintillators for different configurations
and always obtained a similar response within 5\% (see Fig.~\ref{fig:dub}).
We therefore opted for the less expensive Dubna scintillator.
For each test we took the most probable value of a fit to a Landau distribution
as the pe/MIP light yield of our measurement. 

Both response uniformity (better than 3\%) and light yield of tiles were found
equal within errors for two kinds of groove path, circular or sigma ($\sigma$),
passing 5 mm from the tile edges and having a radius of 5 cm at the corners.
A spiral $\sigma$-groove was chosen, as 
shown in Fig.~\ref{fig:tile}.
\begin{figure}[htb]
\vspace*{-.25in}
\epsfig{file=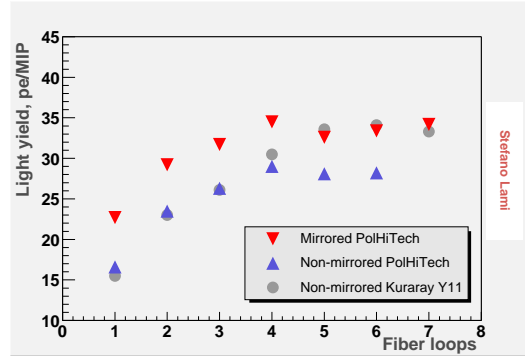,angle=0,width=16.5pc}
\vspace*{-.35in}
\caption{Tile light yield as a function of the number of fiber loops.}
\label{fig:loops}
\vspace*{-.27in}
\end{figure}

We studied two different groove cross sectional shapes.
Originally the baseline design was based on a square cross section, 6 mm deep, 
loaded with 4 or 5 glued loops of WLS fiber. Recently, the design was changed to
a keyhole shape groove
with air coupling, which performed equally well 
%once the length of the fiber inside the tile was extended 
with a 2-loop spiral, as 
seen in Fig.~\ref{fig:tile}.  
For the interested reader, Fig.~\ref{fig:loops} shows the result of our study on
the optimal number of fiber loops into a $\sigma$-groove with a square cross section - 8 mm deep for
this particular test. For Pol.Hi.Tech fibers the light yield reaches a plateau
at N$=$4 loops, after which the gain of a bigger sampling seems to be compensated by the
attenuation of the longer fiber. For a Kuraray Y11(250) fiber, whose
attenuation length is about 50 cm longer than Pol.Hi.Tech or about one extra loop, the plateau is
reached at N$=$5 loops.

\begin{figure}[htb]
\vspace*{-.04in}
\epsfig{file=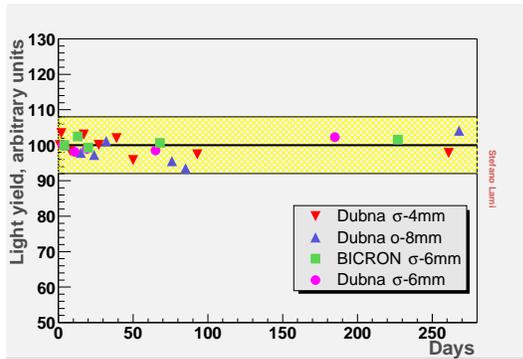,angle=0,width=16.5pc}
\vspace*{-.35in}
\caption{Time response of 4 glued tiles.}
\label{fig:time}
\vspace*{-.22in}
\end{figure}

The concern about glue damaging fibers prompted us to study the response of different
glued tiles over a period of 9 months. No time decay was observed, as shown in Fig.~\ref{fig:time}.
However, to reduce assembly time and remove any concern about gluing,
we decided to use a keyhole groove (1.8 mm deep) where no glue is needed to hold
the fiber.
\begin{figure}[htb]
\vspace*{-.08in}
\epsfig{file=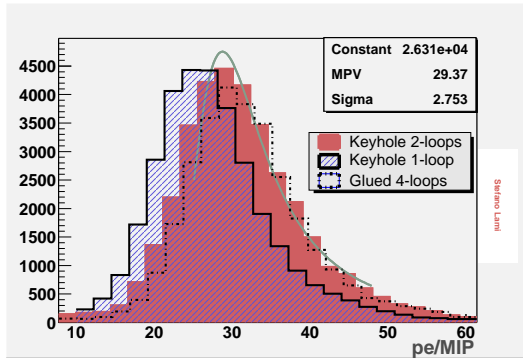,angle=0,width=16.5pc}
\vspace*{-.35in}
\caption{Comparison of alternative methods of embedding a WLS fiber into the tile.}
\label{fig:keyhole}
\vspace*{-.3in}
\end{figure}
In Fig.~\ref{fig:keyhole} the light yield from a 2-loop spiral keyhole groove
with air coupling (solid histogram and fit) is compared to the light yield from
4 glued fiber loops in a 6 mm deep square cross sectional groove (dash-dotted histogram). 
Both methods give a similar result, $\sim$15\% more light than 1 loop of keyhole groove (dashed histogram). 
In this test Tyvek paper and a mirrored Pol.Hi.Tech fiber were used.
The keyhole cross section, with a diameter of 1.2 mm, seems to serve as an extra cladding to better trap
the light lost by the fiber. The spiral is limited to 2 loops, with outer (inner) radius of 5 (4.5) cm,
to mantain the fiber curvature above a radius of 4 cm, critical for fiber stress.
With this design we obtained 38 pe/MIP in the best configuration with air coupling (test 7 in
Tab.~\ref{tab:last}).
\begin{table*}[htb]
\vspace*{-.15in}
\renewcommand{\arraystretch}{0.9}
\caption{Number of pe/MIP for different WLS 
fibers into a 2-loop spiral keyhole groove.}
\vspace*{-.01in}
\label{table:1}
\newcommand{\m}{\hphantom{$-$}}
\newcommand{\cc}[1]{\multicolumn{1}{c}{#1}}
\renewcommand{\tabcolsep}{2pc} % enlarge column spacing
\renewcommand{\arraystretch}{0.92} % enlarge line spacing
\begin{tabular}{@{}cllll}
\hline
 Test &  \cc{Fiber into spiral } & \cc{Mirror} & \cc{Reflector} & \cc{pe/MIP} \\ \hline
  1 &  Pol.Hi.Tech       & No  & Tyvek &  \m24.1   \\
  2 &  Pol.Hi.Tech       & Yes & Tyvek &  \m29.4   \\
  3 &  Kuraray Y11(250)  & No  & 3M &  \m28.1   \\
  4 &  Kuraray Y11(250)  & Yes  & Tyvek &  \m30.5   \\
  5 &  Kuraray Y11(350)  & No  & Tyvek &  \m28.7   \\
  6 &  Kuraray Y11(350)  & No  & 3M &  \m32.9   \\
  7 &  Kuraray Y11(350)  & Yes & 3M &  \m38.2   \\
  8 &  Kuraray Y11(350)+Grease  & Yes & 3M &  \m43.9  \\
\hline
\end{tabular}
  \label{tab:last}
\vspace*{-.2in}
\end{table*}

The study of the best configuration with a  2-loop spiral keyhole groove
in a Dubna tile is summarized in Tab.~\ref{tab:last}.
We obtained a similar light yield with either Pol.Hi.Tech or Kuraray Y11(250)
fibers, but Kuraray Y11(350) were about 17\% better (test 6 vs. test 3).
The 3M reflector increased the light yield by 15\% relative to Tyvek and provided
good uniformity.
Aluminizing the end of the fiber left inside the tile makes an important
difference, $\sim$20\% more light in the  2-loop spiral.
While test 7 is our final choice, we also tested that filling the keyhole
groove with grease would increase the light yield to 44 pe/MIP, 
$\sim$15\% more than air coupling.

Given the narrow width of 5 cm for the CCR tile, a mirrored Kuraray Y11(350)
fiber will be loaded into a straight groove. We obtained 7 pe/MIP (the tile is
5 mm thick), which is acceptable as we will measure electromagnetic
showers in the central calorimeter azimuthal cracks.

%\vspace*{-.05in}
\section{CONCLUSIONS}
The CDF experiment is undergoing an upgrade of the central calorimeter
in order to mantain and extend the Run I capabilities.
Design choices have been made to minimize cost and technical risk.
The light yield obtained exceeds the original design requirements.

At the time of writing, the assembly of the final CPR modules is under way and preliminary
tests show an average light yield of 16 pe/MIP at the PMT (after the whole
optical chain, including 5 m of clear fiber), consistent with a light
yield of $\sim$ 40 pe/MIP at the tile exit.
The new CPR and CCR detectors are to be installed on the front
face of the central calorimeter during Fall 2004.
%
%\vspace*{-.06in}

\vspace*{-.5in}
\end{document}